\documentclass[conference]{IEEEtran}
\IEEEoverridecommandlockouts
\usepackage{cite}
\usepackage[colorlinks,linkcolor=blue,citecolor=blue]{hyperref}

\usepackage{amsmath,amsthm,amsfonts,amssymb,amscd}
\usepackage{mathrsfs}
\usepackage{setspace}
\usepackage{latexsym}
\usepackage{graphicx}     
\usepackage{subfig}
\usepackage{float}
\usepackage{listings}
\usepackage{amsmath,amssymb,amsfonts}
\usepackage{graphicx}
\usepackage{textcomp}
\usepackage{xcolor}
\usepackage{algorithm}
\usepackage{algorithmicx}
\usepackage{algpseudocode}
\usepackage[top=3.2cm,bottom=3.8cm,left=3cm,right=2cm]{geometry}  
\theoremstyle{definition}

\numberwithin{equation}{section}
\def\BibTeX{{\rm B\kern-.05em{\sc i\kern-.025em b}\kern-.08em
    T\kern-.1667em\lower.7ex\hbox{E}\kern-.125emX}}
\begin{document}

\title{A Mathematical Theory of Hyper-simplex Fractal Network for Blockchain: Part I\\
\thanks{*Corresponding Author: Hao Xu (hxu@tongji.edu.cn). \\The list of authors may be adjusted. To make any changes, such as adding or removing an author, please contact the corresponding author with the contributions you wish to add or remove from the work. The preprint's authorship will be updated accordingly.
We are actively seeking collaborations and opportunities to develop this research further. We welcome researchers, institutions, and industry partners interested in exploring the applications and extensions of our Hyper-simplex fractal network theory for blockchain and other complex network systems. If you are interested in collaborating or have ideas for potential applications, please contact the corresponding author.
}
}

\author{Kaiwen Yang, Hao Xu*, Yunqing Sun, Jiacheng Qian, Zihan Zhou,\\ Xiaoshuai Zhang, Erwu Liu, Lei Zhang and Chih-Lin I
}

\maketitle

\begin{abstract}
Blockchain technology holds promise for Web 3.0, but scalability remains a critical challenge. Here, we present a mathematical theory for a novel blockchain network topology based on fractal N-dimensional simplexes. This Hyper-simplex fractal network folds one-dimensional data blocks into geometric shapes, reflecting both underlying and overlaying network connectivities. Our approach offers near-infinite scalability, accommodating trillions of nodes while maintaining efficiency.

We derive the mathematical foundations for generating and describing these network topologies, proving key properties such as node count, connectivity patterns, and fractal dimension. The resulting structure facilitates a hierarchical consensus mechanism and enables deterministic address mapping for rapid routing.
This theoretical framework lays the groundwork for next-generation blockchain architectures, potentially revolutionizing large-scale decentralized systems. 
The Part I work was conducted between March and September 2024. 
\end{abstract}


\section{INTRODUCTION}

Web 3.0 aims to establish a decentralized universal digital identity system, representing the next stage of the World Wide Web. Establishing a decentralized network capable of coordinating trillions of end users' devices is a key requirement for the realization of Web 3.0. However, existing web networking protocols, e.g., IPv4 and IPv6, use Domain Name Servers (DNS) to treat the address with local and global routing based on routing tables within the domain, posing a fundamental challenge to the decentralization on the global scale for both wired and wireless network\cite{jin2021dnsonchain}.

Due to its inherent advantages of immutability, security, and trustworthiness, blockchain technology is widely regarded as possessing the potential to serve as the foundational infrastructure for the realization of Web 3.0\cite{miraz2018applications,bambacht2022web3}. However, despite the promising potential of blockchain technology, there exist two significant challenges that must be addressed for successful real-world implementation. The first challenge pertains to the inherent limitations of the blockchain architecture. Traditionally, blockchain is only considered as a one-dimensional data structure and has limited reflection of the actual  world\cite{li2021blockchain}, although there have been some structural innovations in public networks, such as sidechains\cite{singh2020sidechain}, directed acyclic graphs (DAG)\cite{cao2019internet} and Layer 2 networks\cite{sguanci2021layer}, these solutions are predominantly non-deterministic in nature, resulting in low throughput and slow consensus formation, and some carefully designed deterministic permissioned blockchain networks, such as hyperledger\cite{androulaki2018hyperledger}, R3\cite{guo2016blockchain}and Consensys\cite{mazzoni2022performance}, suffer from limited network extensibility and are unable to be scaled to large-scale applications. The second challenge is closely related to the complexity of communication. Traditional proof-based consensus algorithms, such as Proof-of-Work (PoW)\cite{nakamoto2008bitcoin}, which is well known to require substantial computational resources\cite{KOHLI202379}, while Proof-of-Stake (PoS)\cite{li2017securing} tends to result in monopoly issues, limiting decentralization and fairness \cite{shifferaw2021pos_limitations}. The existing lighter consensus algorithms, such as PBFT\cite{castro1999practical}, Paxos\cite{lamport2019part} and Raft\cite{lamport2019part} don't perform ideal either, because they all exhibit polynomial communication complexity, thus being limited to achieve large-scale extensibility and scalability.

The primary objective of this work is to address those two key challenges identified by proposing a novel blockchain topology structure, along with the design of an efficient consensus algorithm, thus creating a data network that possesses inherent authenticity, spatiotemporal continuity, and large-scale consistency - capabilities that are essential for realizing the Web 3.0 vision. In the realm of topology construction, we draw inspiration from the principles of self-similar fractals\cite{falconer2014fractal,hutchinson1981fractals} , and some applications of self-similar fractals in communication\cite{xu2024web} and computer science\cite{laizet2010numerical,zhang2018towards}. In terms of consensus design, we draw inspiration from the concepts of space-time proof\cite{benet2014ipfs} and multi-layer PBFT\cite{li2020scalable,du2020mbft}, and propose a novel consensus mechanism.

In this paper, we develop a new blockchain topology based on hyper-simplex fractals, which provides near-infinite scalability and a high degree of organization. Additionally, we present a mechanism for network address mapping within the blockchain. Based on the hyper-simplex fractals topology, we design a hierarchical consensus mechanism inspired by multi-layer PBFT\cite{li2020scalable}, which achieves nearly linear computational complexity, albeit with a trade-off in latency. Meanwhile, we develop a blockchain refresh mechanism akin to the Ouroboros, establishing a clear order of block refresh to ensure the consistency of the overall timestamps. A detailed security analysis is also provided. To achieve a true peer-to-peer network, we additionally propose a mechanism termed proof-of-routing, which adaptively adjusts the positions of nodes based on routing capability within the network. This process also facilitates configuration optimization, enhancing the overall efficiency and functionality of the network.

\section{RESULTS}

\subsection{Hyper-simplex fractal topology construction}

To understand the shape of a Hyper-simplex fractal, the key is to investigate how a single iteration will change a face.

Obviously, the face, together with the nodes adjacent to it, constructs a $(N-1)-$simplex network. When an iteration occurs, the midpoint of each edge and a distance out of the simplex along the normal direction outside the simplex will generate a new node, which will generate a total of $N$ new nodes per face. Sometimes, the nodes generated by two adjacent faces will appear in the same position, and they need to be treated as different nodes.

\begin{figure}[H]
	\centering
	\includegraphics[width=0.9\linewidth]{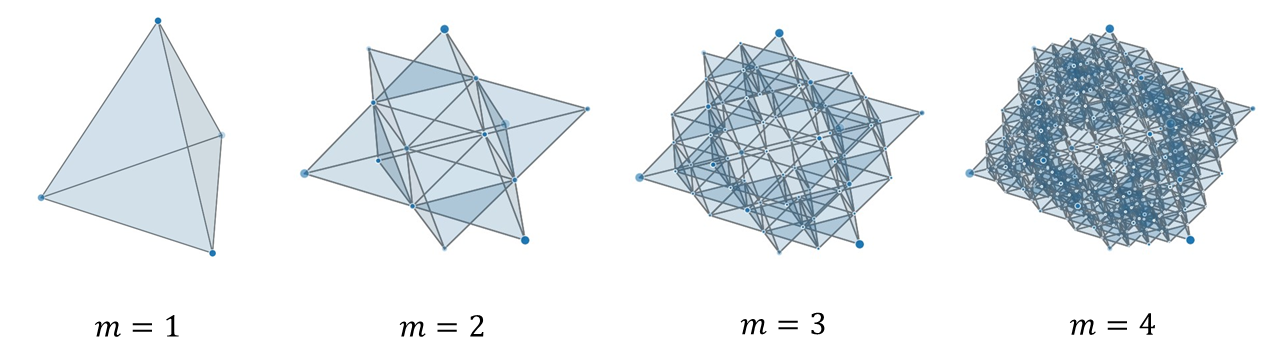}
	\caption{$4$-simplex fractals}
\end{figure}

By utilizing Iterated Function Systems (IFS)\cite{barnsley1985iterated} in mathematics, we can clearly articulate the generation process of fractals and calculate that the number  $V_{N}^m$  of $N$-simplex fractals network nodes after $m$ iterations is approximately

\begin{equation}
	V_{N}^m\approx 2^{m-2}N^m.
\end{equation}

\subsection{Scale-invariant representation, address mapping and routing}

Fractals themselves are geometric structures that exhibit scale invariance, and this property also extends to the representation of nodes within a fractal network. Here, we introduce the subscript pair representation to represent nodes, while address mapping serves as a direct translation of these indexed pairs. 

By utilizing duality, we can alternately and recursively represent each node and face within the fractal network, and the number of subscript pairs in the subscript represents which tier it is generated in. The detailed representation methods can be found in the corresponding METHODS section.

\begin{figure}[h!]
	\centering
	\includegraphics[width=0.5\linewidth]{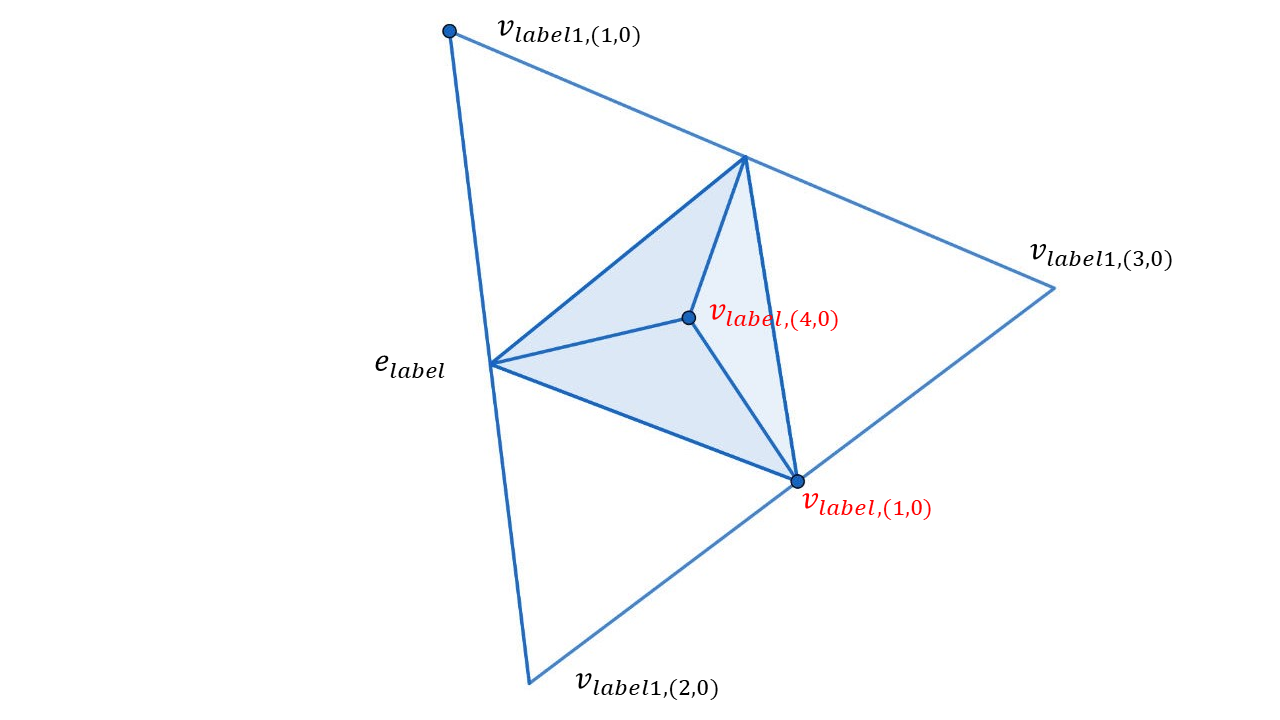}
	\caption{Node's recursively representation}
\end{figure}

\begin{figure}[h!]
	\centering
	\includegraphics[width=0.5\linewidth]{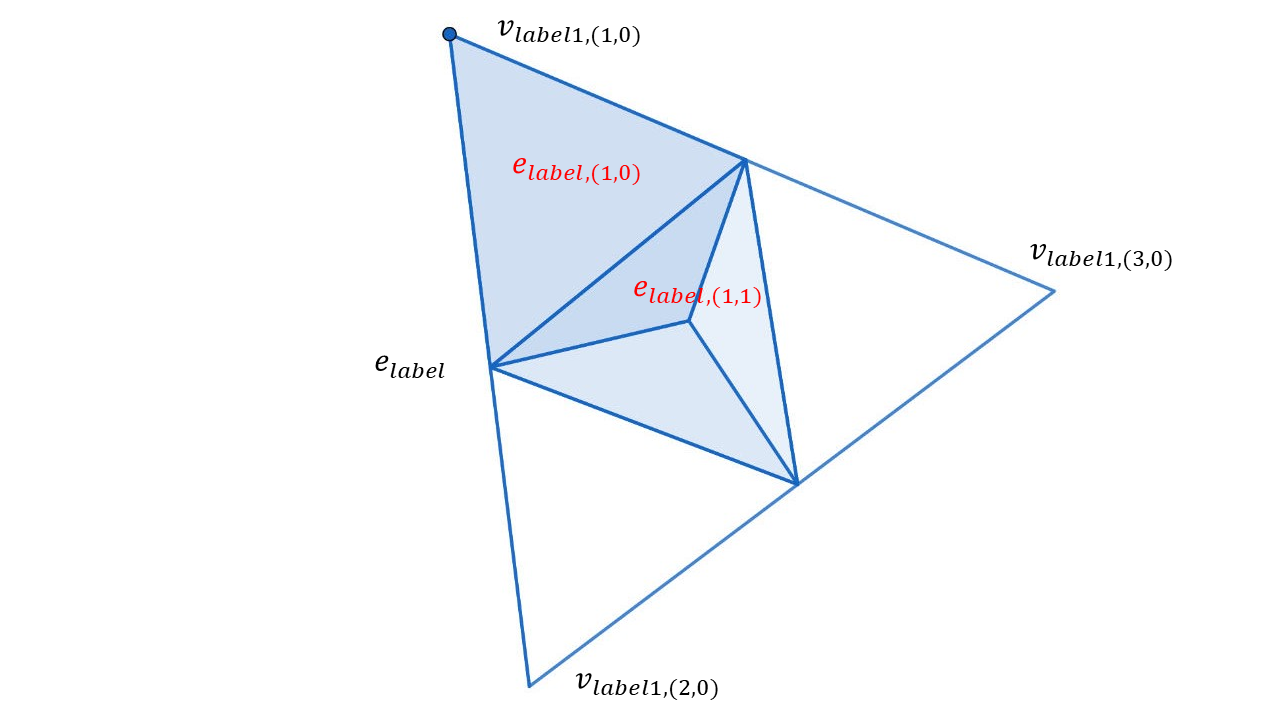}
	\caption{Face's recursively representation}
\end{figure}

	Hyper-simplex fractal network works similarly of label-based routing and switching, e.g.,  Multiprotocol Label Switching (MPLS)\cite{rosen2001multiprotocol}, and take the principles of "Switch where possible, route if necessary". Upon the entry of the Hyper-simplex fractals network edge, the Hyper-simplex fractals network locator is used to create the pseudo-deterministic yet robust paths before dispatching the flow. The label will be generated at the time of connecting to any part of the network, and it will also carry the deterministic routes using the tier number locator.
	
	The detailed label-making process is dependent on the size of the number of top-level nodes $N$ and the tier number $m$. The subscript pair representation of a node exactly reflects the iter position of the node in Hyper-simplex fractals network; for each subscript pair in the subscript pair representation, $\lceil\log_2N\rceil+1$ bits are needed to represent, the first $\lceil\log_2N\rceil$ bits represent the first digit of the subscript pair, which ranges from $1$ to $N$, and the other bit represents the second digit of the subscript pair, which can be $0$ or $1$. To determine the number $N$, an additional $\lceil\log_2N\rceil$ bits are needed. So totally
	 \begin{equation}
	 	m(\lceil\log_2N\rceil+1)+\lceil\log_2N\rceil=(m+1)\lceil\log_2N\rceil+m
	 \end{equation} bits are needed to determine the dimension and iter number of one Hyper-simplex fractal network and to be a tier locator of $K_{N,m}$. When the subscript pair is not long enough, all unset numbers are set to $1$ because the last subscript of an existing node can not be $(k,1),\ 1\le k\le N$. Thus, we can locate the last zero in the tier locator to determine which iteration the node is generated in.
	
	Let $V_{N,m}$ be the node set of $N-$simplex fractal with tier $m$. For example, $v_{(2,0)}$'s tier locator in $V_{3,3}$ is
	$$
	10:010:111:111
	$$ 
where the first 10 means $N=2+1=3$ and there are $3$ sets of numbers behind so $m=3$. Similarly, $v_{(1,0),(2,1),(3,0)}$'s tier locator in $V_{3,3}$ is
	$$
	10:000:011:100
	$$
	
	If a node's tier locator is
	$$
	11:000:011:100:111
	$$
	It will be $v_{(1,0),(2,1),(3,0)}$ in $V_{4,4}$ because the first $11$ means $N=3+1=4$  and there are $4$ sets of numbers behind so $m=4$, and the last zero in its tier locator appears in the fourth set of numbers means this node is generated in the third iteration.

\subsection{Consensus tree and hierarchical consensus mechanism}

We can consider all the nodes of the same small simplex as a whole, and with the help of the concept of mathematical quotient mapping\cite{bellettini2004quotient}, reorganize the nodes on a Hyper-simplex fractal network. All the nodes of the same small simplex will form a consensus node, and all the consensus nodes will be organized as a tree, which we call a consensus tree. For a $N$-simplex fractal, there is only one consensus node in the lowest layer, while the $k$-th layer contains a total of $N(2N-2)^{k-2}$ consensus nodes. The consensus node at the lowest layer directly leads $N$ consensus nodes, whereas each of the other non-top-layer consensus nodes directly leads $2(N-1)$ consensus nodes.

A Multi-Layer PBFT consensus is operated on the Hyper-simplex fractal network. In one consensus node, a traditional PBFT\cite{castro1999practical} consensus is operated, and in this process, they will also vote a leader node to reply to the result to each node in its father consensus node, forming the downward propagation of consensus, and repeat this process until it reaches the lowest consensus node. Similarly, once the father consensus node has formed a consensus, its leader node will send the message to each node of all of its directly direct successor nodes except for the message source.

\begin{figure}[h!]
	\centering
	\includegraphics[width=0.85\linewidth]{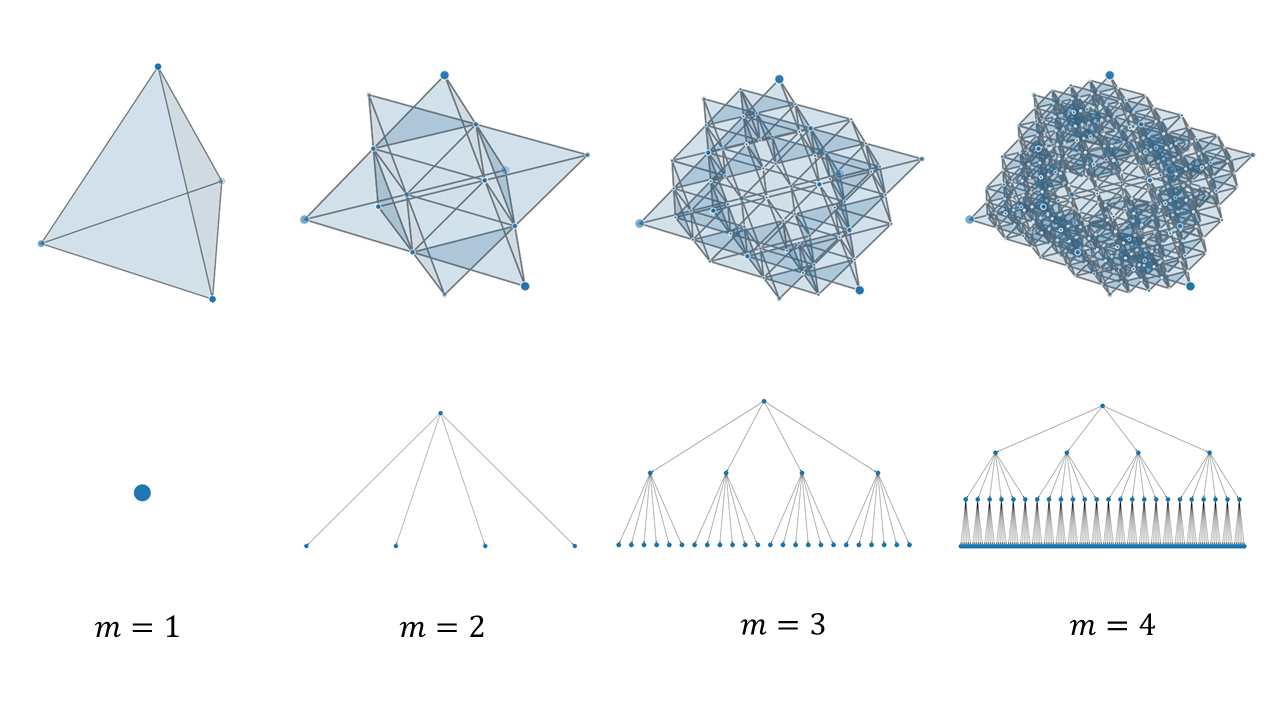}
	\caption{Illustration of Consensus Tree with Number $m$ of iterations }
\end{figure}

\begin{figure}[h!]
	\centering
	\includegraphics[width=0.85\linewidth]{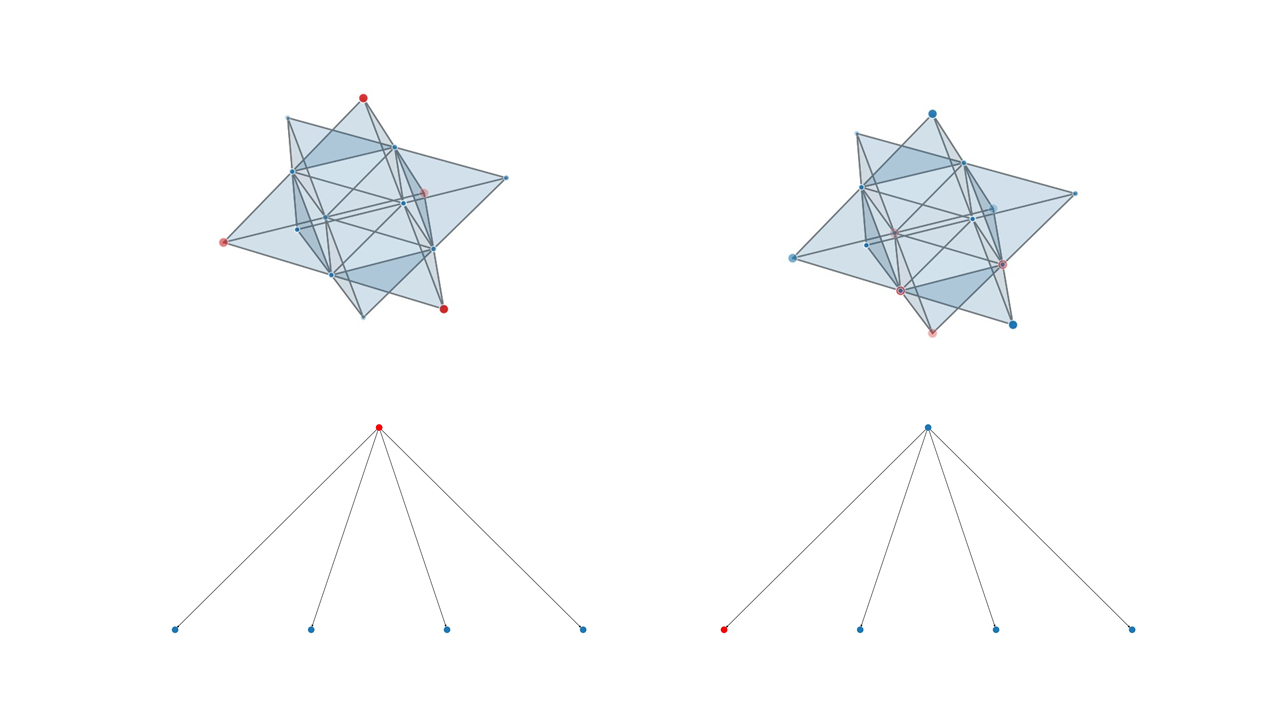}
	\caption{Schematic diagram of the correspondence between the consensus node and the original node}
\end{figure}

\subsection{Communication complexity analysis and latency analysis}

For a given fractal network, when populating nodes, we aim to fill from the lower layer to the upper layer and ensure that the nodes at the highest layer are approximately uniformly distributed among the various consensus nodes. This setup helps guarantee that the network operates at maximum efficiency.

Calculations show that the communication complexity $C$ satisfies
\begin{equation}
	C\approx V^{1+\frac{\log_2 2N}{\log_2 V}},
\end{equation}
where $V$ represents the total number of nodes. This is a highly favorable result; when $V$ is very large, the relationship between $C$ and $V$ approaches linearity. For instance, when $N=10$ and $V=10^{11}$, the communication complexity $C$ only about $10^{11}$, far superior to the traditional PBFT's $10^{20}$. 

But every good thing has its trade-off. The increased conﬁrmation delay comes with a significant reduction in communication complexity. Assuming that each layer
takes average $t_{ave}$ to reach consensus and propagate to the adjacent layer, then the average propagation delay $D$ satisfies
\begin{equation}
	D\approx \frac{\log_2 V}{\log_2 2N}t_{ave}.
\end{equation}

In reality, networks, although we can use parallel routes and distribute different information through different paths to reduce $t_{ave}$, the time required to reach consensus will still significantly increase as $V$ grows. Therefore, it is important to strike a balance between communication complexity and conﬁrmation delay when designing the Hyper-simplex fractal network network.

\begin{figure}[h!]
	\centering
	\includegraphics[width=0.85\linewidth]{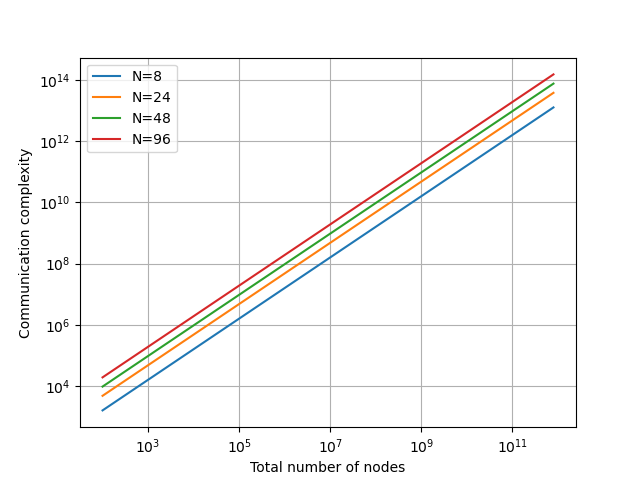}
	\caption{Communications Complexity of N-Simplex Network with $N$ = 8/24/48/96}
\end{figure}

\begin{figure}[h!]
	\centering
	\includegraphics[width=0.85\linewidth]{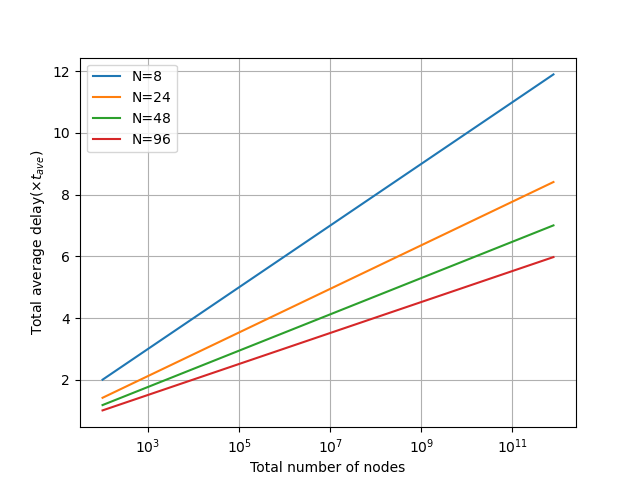}
	\caption{Communications latency (normalized) of N-Simplex Network with $N$ = 8/24/48/96}
\end{figure}

\subsection{The ouroboros-like update order}

The hierarchical consensus mechanism mentioned above is a parallel consensus mechanism that cannot ensure network synchronization. To unify timestamps and ensure the network operates in an orderly manner, it is also necessary to design the update order of the space-time blockchain, which resembles an Ouroboros-like update order.

By constructing a loop that traverses all nodes of the Hyper-simplex fractal network, where nodes within the same consensus subtree are adjacent in the traversal order, we can propose a feasible update sequence for the space-time blockchain. The design of this traversal loop is somewhat like a depth-first search. In fact, after introducing the concept of consensus trees, this traversal loop is indeed a depth-first search on consensus trees.

\begin{figure}[h!]
	\centering
	\includegraphics[width=0.85\linewidth]{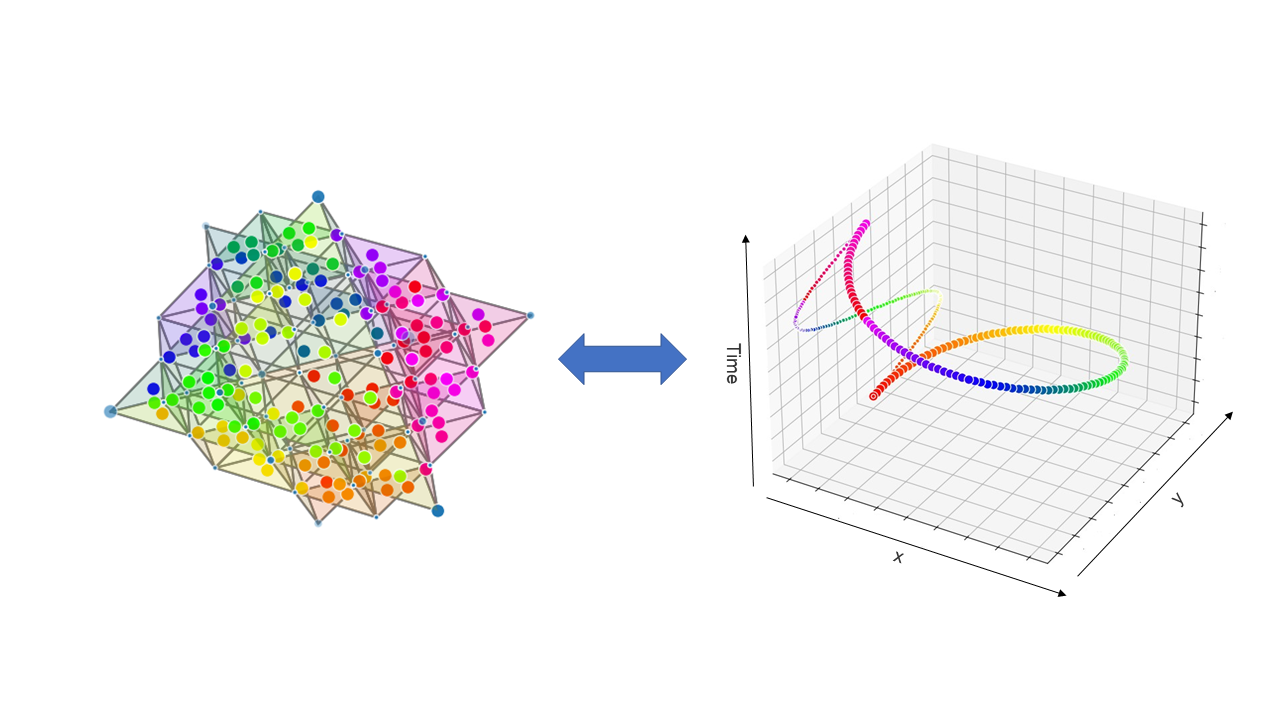}
	\caption{Node traversal order (the color of each face is the average color of its vertices).}
\end{figure}

\begin{figure}[h!]
	\centering
	\includegraphics[width=0.85\linewidth]{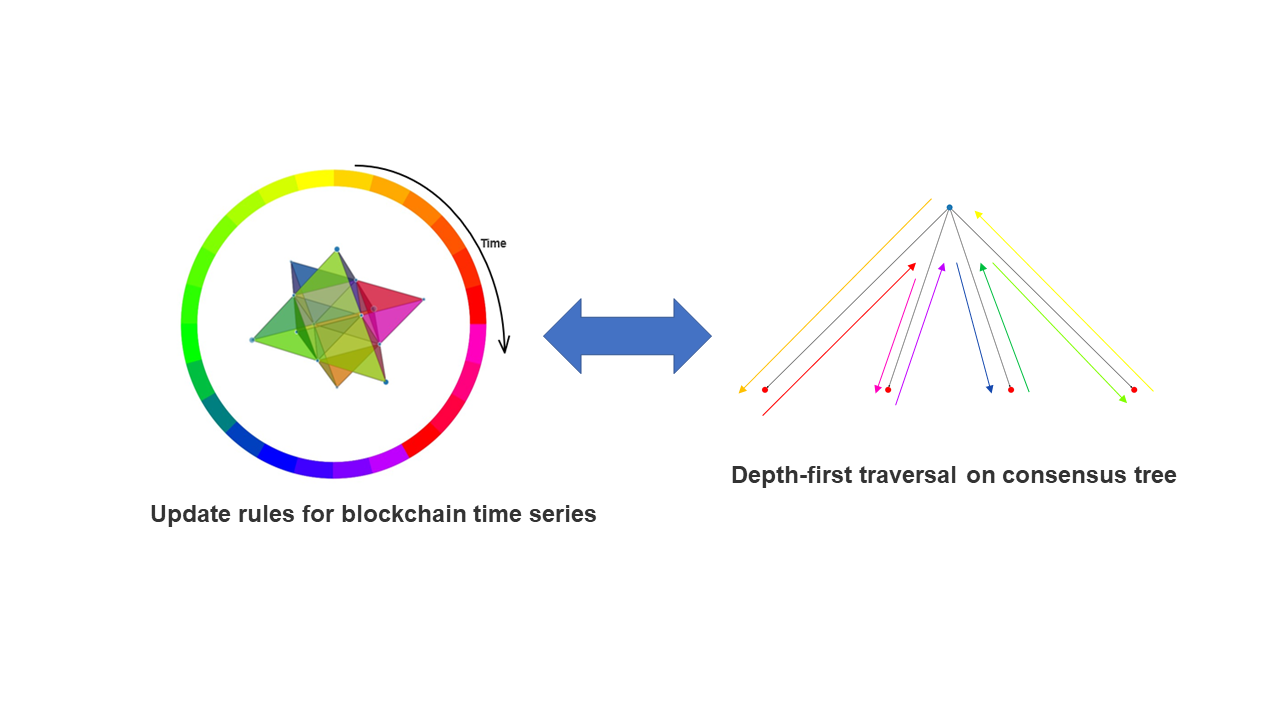}
	\caption{The traversal order corresponds to depth-first search on the consensus tree.}
\end{figure}

In practical applications, it is possible for certain nodes to be empty or faulty. In such cases, during the update process of the space-time blockchain, these nodes can simply be skipped temporarily. When the number of nodes is very large, the waiting time for update intervals may be lengthy. In such cases, it is necessary to implement some optimization strategies, like region partitioning, loop nesting, etc, even at the cost of sacrificing some synchronization.

\subsection{Security analysis and fault tolerance evaluation}

There are two main malicious attack models in distributed systems and blockchain: the faulty probability determined (FPD) model and the faulty number determined (FND) model\cite{satyanarayanan1989integrating,anderson2020security}, used when the probability of every single faulty node is ﬁxed and the number of faulty nodes in the system is ﬁxed respectively. For FPD, the failure probability of each node is independent, and the number of failed nodes within a certain range follows a binomial distribution. In contrast, for FNP, since the total number of failed nodes is fixed, the failure probability of each node is dependent on whether other nodes have failed, and the number of failed nodes within a certain range follows a hypergeometric distribution. However, when the number of nodes is large, the hypergeometric distribution approximates the binomial distribution\cite{rice2007mathematical}, so the results for FND are similar to those for FPD. Our simulation experiments also support this observation. Since the Hyper-simplex fractal network is primarily used in large-scale scenarios, we will mainly focus on safety analysis for FPD in this section.

Let the failure probability of a single node be denoted as $P_f$ and the consensus success rate of $N$-simplex fractals network nodes after $m$ iterations as $P_N^m$. Under our update rules, the blockchain can operate normally if and only if its lowest layer consensus node functions properly. For each consensus node, it can operate normally if and only if there are no more than $\left \lfloor \frac{N}{3} \right \rfloor$  failed nodes within it, and less than half of its child consensus nodes have failed. Thus, we can recursively derive the calculations:

The failure probability of the consensus nodes at layer $m$ is
\begin{equation}
	P_m=1-\sum_{i=0}^{\left \lfloor \frac{N}{3} \right \rfloor}C_N^i(1-P_f)^{N-i}P_f^i.
\end{equation}
and the recursive calculation formula is
\begin{equation}
\begin{aligned}
P_k = 1 &- \sum_{i=0}^{N-2}C_{2N-2}^i(1-P_{k+1})^{2N-2-i}P_{k+1}^i + P_m \\
&- P_m(1-\sum_{i=0}^{N-2}C_{2N-2}^i(1-P_{k+1})^{2N-2-i}P_{k+1}^i), \\
&\quad 1<k<m.
\end{aligned}
\end{equation}
Thus eventually
\begin{equation}
\begin{aligned}
P_N^m = 1 &- \sum_{i=0}^{\left \lfloor \frac{N}{2}\right \rfloor}C_N^i(1-P_2)^{N-i}P_2^i \\
&+ P_m - P_m\left(1-\sum_{i=0}^{\left \lfloor \frac{N}{2}\right \rfloor}C_N^i(1-P_2)^{N-i}P_2^i\right).
\end{aligned}
\end{equation}

\subsection{Dynamic Node Positioning Based on Node Performance}

The fractal network topology inherently creates a hierarchical structure where nodes at different levels face varying demands and responsibilities. We propose a dynamic node positioning system that leverages this characteristic to create a truly peer-to-peer network by adjusting node positions based on their demonstrated overall performance.

In this system, nodes are not permanently fixed to their initial positions within the network. Instead, their placement is fluid and determined by their performance metrics. The fundamental principle is straightforward: nodes that exhibit superior performance are promoted to higher levels within the fractal structure, while those with lesser capabilities may be moved to lower levels or expelled from the consensus group, i.e., losing the vertices.

This approach offers several advantages:

\begin{enumerate}
    \item \textbf{Optimal Resource Utilization:} By positioning high-performance nodes at higher levels where they handle more critical tasks, the network can more efficiently utilize its resources.
    
    \item \textbf{Natural Load Balancing:} As stronger nodes ascend the hierarchy, they naturally take on more responsibilities, distributing the network load more evenly.
    
    \item \textbf{Incentivization for Network Improvement:} Nodes are incentivized to improve their overall performance to gain higher positions in the network, potentially leading to overall network enhancement.
    
    \item \textbf{Adaptive Network Structure:} The network can dynamically adapt to changes in node performance, maintaining optimal efficiency even as individual node capabilities fluctuate over time.
    
    \item \textbf{True Peer-to-Peer Paradigm:} By allowing any node to potentially occupy any position based solely on its performance, this system embodies the essence of a decentralized, peer-to-peer network.
\end{enumerate}

The implementation of this system involves continuous monitoring of various node performance metrics, which may include but are not limited to processing speed, storage capacity, uptime, and network connectivity. Periodically or triggered by significant performance changes, the network initiates a rebalancing process. During this process, high-performing nodes from lower levels may replace underperforming nodes at higher levels, adhering to the fractal topology's structural constraints.

This dynamic positioning mechanism ensures that the network's topology remains optimal in terms of overall efficiency while maintaining its fractal properties. It represents a novel approach to network organization that aligns node hierarchy with node capability, potentially setting a new standard for scalable and efficient blockchain networks.

\section{METHODS}

\subsection{Iterated Function Systems (IFS) and computation}

If we write the node set of $N$-simplex fractal with tier $m$ as $V_{N,m}=\cup_{i\in \text{label}}v_{i}$ and the face set of $N$-simplex fractal with tier $m$ as $E_{N,m}=\cup_{i\in \text{label}}e_{i}$. Define the operators $T_1$, $T_2$ satisfying
\begin{equation}
	\begin{aligned}
	    &T_1(\cup_{i=1}^ne_i)=\cup_{i=1}^nT_1(e_i),\\
		&T_2(\cup_{i=1}^ne_i)=\cup_{i=1}^nT_2(e_i),
	\end{aligned}
\end{equation}
 where $e_i$ is a simplex, and $T_2(e_i)$ means the set of the nodes lying on the midpoint of each edge of $e_i$ or a distance out of $e_i$ along the normal direction outside the simplex, $T_1(e_i)$ means the faces determined by the midpoint of these co node edges combining with the node and the node outside $e_i$ respectively. So $V_{N,m}$ and $E_{N,m}$ can be represented as

\begin{equation}
	\begin{aligned}
	    &E_{N,m}=T_2^{(m-1)}(E_{N,1}),\\
		&V_{N,m}=\cup_{i=1}^mT_1^{(m-1)}(E_{N,1}),
	\end{aligned}
\end{equation}
where $E_{N,1}$ is the face set of regular $N-$simplex and define $T_1^{(0)}(E_{N,1})=V_{E,1}$  as the node set of regular $N-$simplex. So the algorithm of the construction of simplex fractal can be written.

\begin{algorithm}[h] 
	\renewcommand{\algorithmicrequire}{\textbf{Input:}}
	\renewcommand{\algorithmicensure}{\textbf{Output:}}
	\caption{The construction of Hyper-simplex fractal} 
	\label{alg::conjugateGradient} 
	\begin{algorithmic}[1] 
		\Require 
		The tier number of simplex fractal $m$, the dimension of Hyper-simplex fractal $(N-1)$.
		\Ensure 
		The node set of simplex fractal $V$, the face set of simplex fractal $E$.
		\State $ V $ ← Initialize $V$ according to the vertices of a regular $N$-simplex with the center of mass at the origin;
		\State $ E $ ← Initialize $E$ according to the faces of a regular $N$-simplex with the center of mass at the origin;
		\State $height$← The height of the regular $N$-simplex with the center of mass at the origin;
		\State $n$←1
		\While {$n<m$}
		\State$E_1$← empty set
		\For {face \textbf{in} $E$}
		\State  $V$ ← $V\cup 
  \{ \mbox{The midpoint of each face edge}\}$
		\State  $en$←Unit outer normal vector of the face corresponding to the face
		\State $V$ ← $V\cup \{$The center of mass of the face corresponding to the face$+height\cdot\frac{en}{2^{n}}\}$
		\State $E_1$←$E_1\cup \{\mbox{The new $2(N-1)$ new faces}\}$
		\EndFor
		\State $E$	← $E_1$
		\State $n$	← $n+1$ 
		\EndWhile
		\State\Return $V$,$E$;
	\end{algorithmic} 
\end{algorithm}

Since a total of $N$ new nodes and $2(N-1)$ new faces will be generated per face in each iteration, the following recursive formula holds,

\begin{equation}
	\begin{aligned}
	    	&|V_{N,m}|=|V_{N,m-1}|+N|E_{N,m-1}|,\\
		&|E_{N,m}|=2(N-1)|E_{N,m-1}|,\\
		&|V_{N,1}|=|E_{N,1}|=N.
	\end{aligned}
\end{equation}

Solve the recurrence formula to get
\begin{equation}
	\left\{\begin{aligned}
		&|V_{N,1}|=N,|V_{N,m}|=N+N^2\frac{1-(2N-2)^{m-1}}{1-(2N-2)},\\
		&|E_{N,m}|=N*(2N-2)^{m-1}.
	\end{aligned}\right.
\end{equation}
and there is an approximate expression
\begin{equation}
	|V_{N,m}|\approx 2^{m-2}N^m.
\end{equation}

	\begin{figure}[htbp]
	\centering
	\begin{minipage}[b]{0.24\textwidth}
		\includegraphics[width=\textwidth]{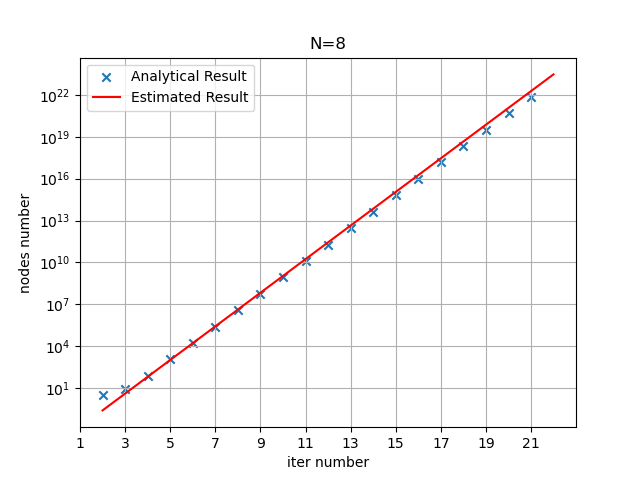}
		\caption{N=8}
		\label{image1}
	\end{minipage}
	\hfill
	\begin{minipage}[b]{0.24\textwidth}
		\includegraphics[width=\textwidth]{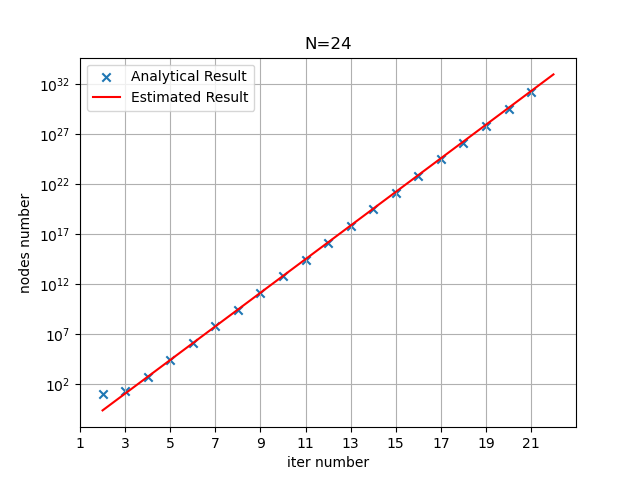}
		\caption{N=24}
		\label{image2}
	\end{minipage}
	\hfill
	\begin{minipage}[b]{0.24\textwidth}
		\includegraphics[width=\textwidth]{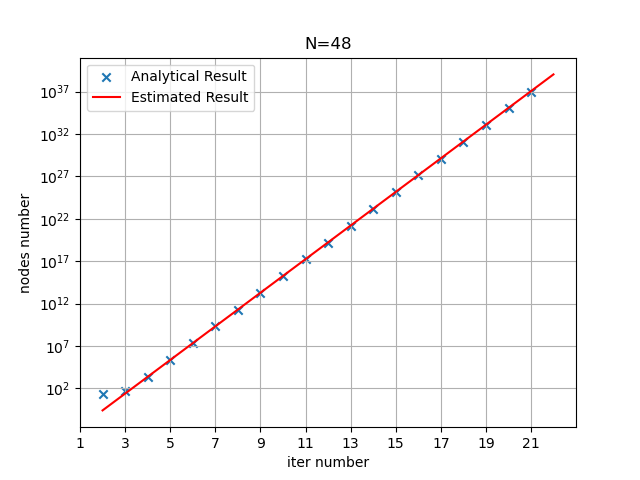}
		\caption{N=48}
		\label{image3}
	\end{minipage}
	\hfill
	\begin{minipage}[b]{0.24\textwidth}
		\includegraphics[width=\textwidth]{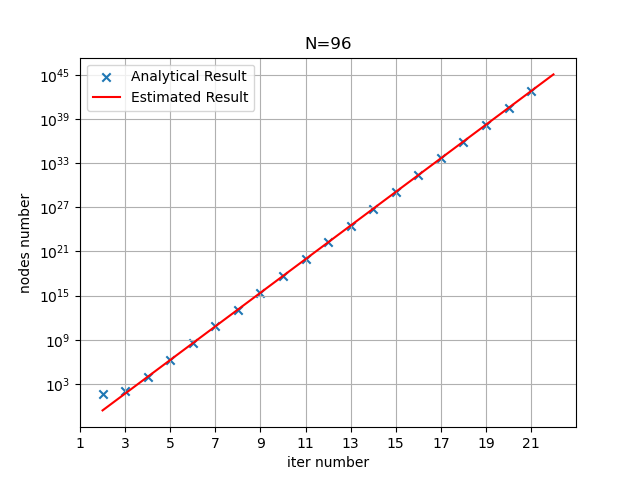}
		\caption{N=96}
		\label{image4}
	\end{minipage}
	\caption{Nodes number vs number of iterations}
\end{figure}

\subsection{Recursive representation of nodes}

\textbf{1) Elements in $V_{N,1}$ and $E_{N,1}$:}

The elements in $V_{N,1}$ can be represented as $v_{(1,0)},v_{(2,0)},...,v_{(N,0)}$, and the face lying on the opposite side of $v_{(k,0)}$ in the simplex is represented as $e_{(k,0)}$.

\textbf{2) Elements in $V_{N,2}-V_{N,1}$ and $E_{N,2}$:}

Record the node generated by $e_{(k,0)}$ and lying outside the $(N-2)$-dimensional simplex determined by $e_{(k,0)}$ as $v_{(k,0),(k,0)}$, and the one lying inside the $(N-2)$-dimensional simplex and on the opposite side of $v_{(l,0)}(l\neq k)$ as $v_{(k.0),(l,0)}$.

Then consider the faces generated by $e_{(k,0)}$. When it is determined by $v_{(l,0)}$ and the midpoint of all edges with $v_{(l,0)}$ as an edge node, it can be represented as $e_{(k,0),(l,0)}$, and when it is determined by $v_{(k,0),(k,0)}$ and the midpoint of all edges with $v_{(l,0)}$ as an edge node, it is represented as $e_{(k,0),(l,1)}$.

\textbf{3) Elements in $V_{N,m}-V_{N,m-1}$ and $E_{N,m}(m>2)$:}

The following recursively represents the elements generated from $e_{label}$.

Firstly, we consider the nodes. Naturally, the one lying inside the $(N-2)$-dimensional simplex and on the opposite side of $v_{label^\prime }$ can be represented as $v_{label,(l,0)}$, where $l$ is the first digit of the last pair of $label^\prime $. As for the node lying outside the $(N-2)$-dimensional simplex determined by $e_{label}$, the situation will be more complicated. There are $(N-2)$ different vertex nodes of the $(N-2)$-dimensional simplex determined by $e_{label}$, and the first digit of their last subscript pairs are different from each other. So there exists a unique positive integer $k$ that is less than or equal to $N$, which is different from the first digit in the last subscript pairs of those $(N-2)$ vertex nodes, and the node to be represented can be represented as $v_{label,(k,0)}$.

Considering faces, the face determined by $v_{label^\prime}$ and the midpoint of all edges with $v_{label^\prime}$ as an edge node can be represented as $e_{label,(l,1)}$, and the face determined by $v_{label,(k,0)}$ and the midpoint of all edges with $v_{label^\prime}$ as an edge node can be represented as $e_{label,(l,2)}$.

Note that set $Label_{N,m}$ is all subscript pairs with $m$ pairs satisfying the first pair can be represented as $(i, 0), i=1,...,N$, followed by $m-1$ subscript pairs $(i, j), i=1,..,N; j=0,1$ with the first digit of every subscript pair being different from the previous, so that $|Label_{N,m}|=N*(2N-2)^{m-1}$ and $E_{N,m}$, $V_{N,m}$ can be represented as
\begin{equation}
    \begin{aligned}
E_{N,m} &= \{e_{label}, label \in Label_{N,m}\}, \\
V_{N,m+1} - V_{N,m} &= \{v_{label,(i,0)}, label \in Label_{N,m}, \\
                    &\quad i=1,\ldots,N\}, \\
V_{N,m} &= \bigcup_{i=2}^{m}(V_{N,i}-V_{N,i-1}) \\
        &\quad \cup \{v_{(i,0)}, i=1,\ldots,N\}.
\end{aligned}
\end{equation}

\subsection{Organization of consensus tree}

Define the mapping $\pi$ on $V_{N,m}$ as 
\begin{equation}
	\begin{aligned}
		&\pi(v_{(i,0)})=\Theta_{(0,0)},
		\\
		&\pi(v_{label,(i,0)}) =\Theta_{label}.
	\end{aligned}
\end{equation}
where
\begin{equation}
	\begin{aligned}
		&\Theta_{(0,0)}=[\cup_{i=1}^Nv_{(i,0)}],
		\\
		&\Theta_{label}=[\cup_{i=1}^Nv_{label,(i,0)}].
	\end{aligned}
\end{equation}
are equivalent classes generated by $\pi$ and are named as consensus nodes.

The subscript pair representation is still used to represent the consensus node, thus naturally forming a hierarchical relationship. Examine this hierarchical relationship and all the consensus nodes can be organized as a consensus tree.

We can write the consensus tree of $N$-simplex fractals network with iter $m$ as $T_{N,m}$. In $T_{N,m}$, there are totally $\frac{|V_{N,m}|}{N}=1+N\frac{1-(2N-2)^{m-1}}{1-(2N-2)}$ consensus nodes, and there are $|E_{N,n-1}|=N*(2N-2)^{n-2}$ consensus nodes in $n$-level($n\ge 2$) consensus layer in the consensus tree.

On the consensus tree, the consensus node $\Theta_{\text{label1}}$ is a subordinate node of $\Theta_{\text{label2}}$ if and only if $\text{label2}$ is the prefix of $\text{label1}$, which means $e_{\text{label1}}$ is generated from $e_{\text{label2}}$ by iteration. The top-level consensus node has $N$ direct subordinate nodes, and this subordinate relationship is
\begin{equation}
	\Theta_{(0,0)}\to\{\cup_{i=1}^{N}\Theta_{(i,0)}\}.
\end{equation}

Since each face generates $2(N-1)$ faces in one iteration, every consensus node has $2N-2$ direct subordinate nodes in $T_{N,m}$ except these in $k$-level, and these relationships are
\begin{equation}
	\Theta_{\text{label}}\to \{\cup_{i=1}^2\cup_{j=1}^N\Theta_{\text{label},(j,i)}/\cup_{i=1}^2\Theta_{\text{label},(k,i)}\},
\end{equation}
where $k$ is the first digit of the last pair of $\text{label}$.

\begin{figure}[h]
	\centering
	\includegraphics[width=0.85\linewidth]{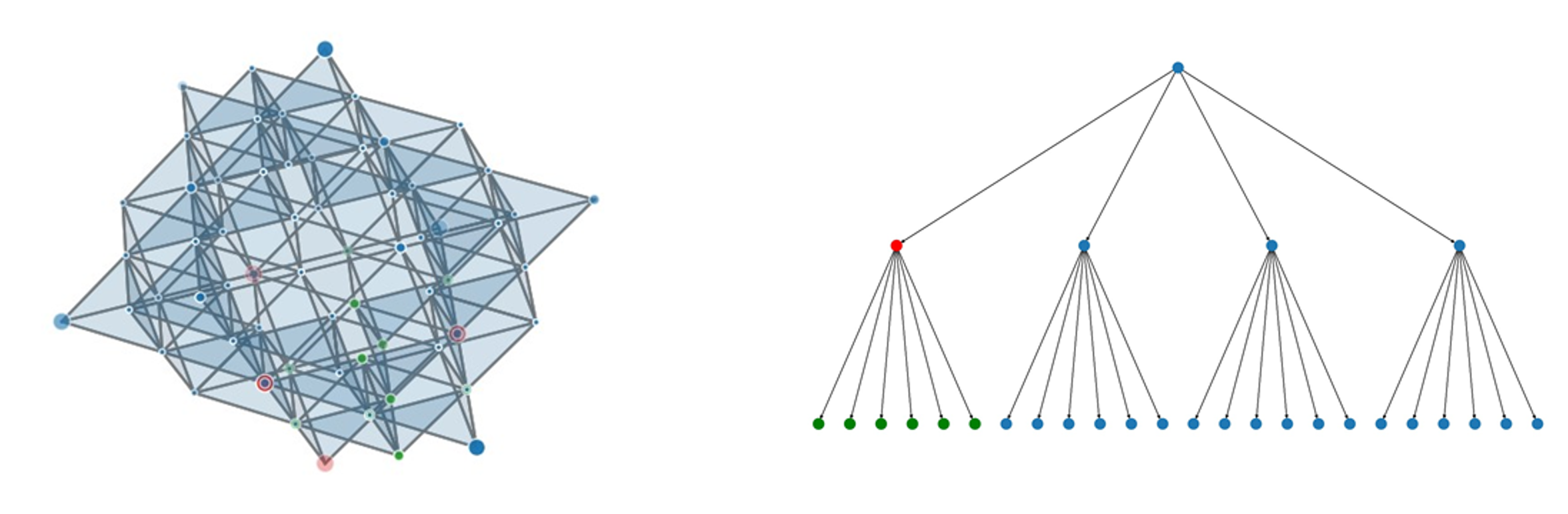}
	\caption{Schematic diagram of subordinate relationships between consensus nodes}
\end{figure}

\subsection{Calculation of communication complexity and propagation delay}

For a filled $N$-simplex fractal network with iter $m$, the
communication complexity $C$ to reach consensus is

\begin{equation}
	C=(N+1)(N+N^2\frac{1-(2N-2)^{m-1}}{1-(2N-2)}),
\end{equation}
where $N(N+N^2\frac{1-(2N-2)^{m-1}}{1-(2N-2)})$ is intra-layer communication complexity and $N+N^2\frac{1-(2N-2)^{m-1}}{1-(2N-2)}$ is the inter-layer communication complexity.

Under general conditions, take $m$ to be the least number satisfying
\begin{equation}
	V<N+N^2\frac{1-(2N-2)^{m-1}}{1-(2N-2)}.
\end{equation}
Define
\begin{equation}
	r= [V-(N+N^2\frac{1-(2N-2)^{m-2}}{1-(2N-2)})]/(N*(2N-2)^{m-2}), 
\end{equation}
then the communication complexity $C$ will be
\begin{equation}
\begin{split}
C = & N^2 + N^3\frac{1-(2N-2)^{m-2}}{1-(2N-2)} \\
    & + [\left \lfloor r \right \rfloor ^2(r-\left \lfloor r \right \rfloor) \\
    & + (\left \lfloor r \right \rfloor +1)^2(\left \lfloor r \right \rfloor+1-r)] \\
    & \times (N*(2N-2)^{m-2}) + V,
\end{split}
\end{equation}

where $N^2+N^3\frac{1-(2N-2)^{m-2}}{1-(2N-2)}$ is the intra-layer communication complexity of the previous $m-1$ fully populated layers, $[\left \lfloor r \right \rfloor ^2(r-\left \lfloor r \right \rfloor )+(\left \lfloor r \right \rfloor +1)^2(\left \lfloor r \right \rfloor+1-r)]*(N*(2N-2)^{m-2})$ is the intra-layer communication complexity of the highest layer, and $V$ is the inter-layer communication complexity.

When $N$ is ﬁxed and nodes are organized by optimal allocation, the relationship between communication complexity $C$ and total node number $V$ can be written as
\begin{equation}
	C\approx V^{1+\frac{\log_2 2N}{\log_2 V}}.
\end{equation}

	\begin{figure}[htbp]
		\centering
		\begin{minipage}[b]{0.24\textwidth}
			\includegraphics[width=\textwidth]{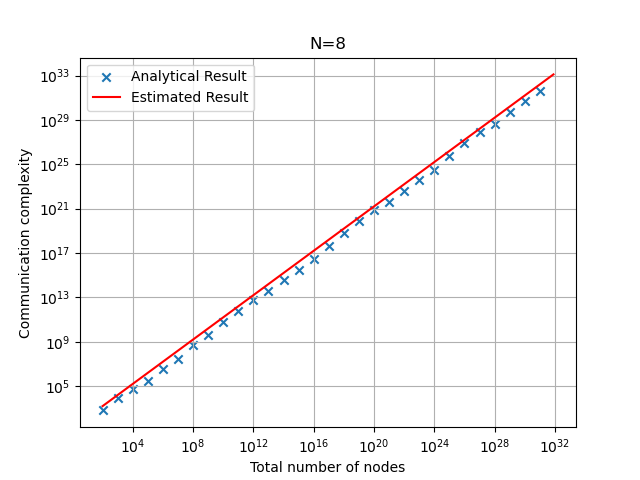}
			\caption{N=8}
			\label{fig:image1}
		\end{minipage}
		\hfill
		\begin{minipage}[b]{0.24\textwidth}
			\includegraphics[width=\textwidth]{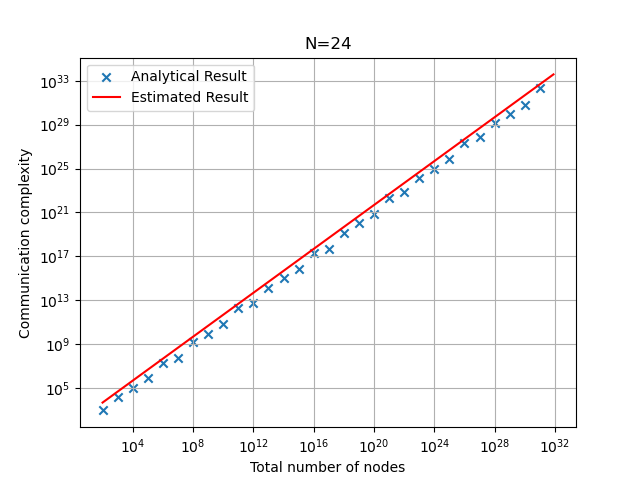}
			\caption{N=24}
			\label{fig:image2}
		\end{minipage}
		\hfill
		\begin{minipage}[b]{0.24\textwidth}
			\includegraphics[width=\textwidth]{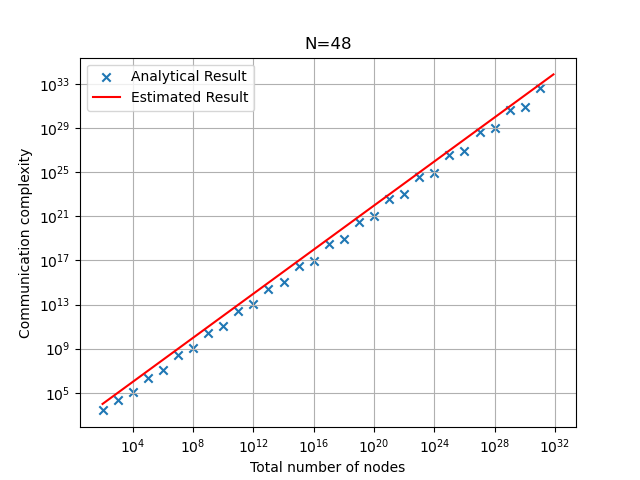}
			\caption{N=48}
			\label{fig:image3}
		\end{minipage}
		\hfill
		\begin{minipage}[b]{0.24\textwidth}
			\includegraphics[width=\textwidth]{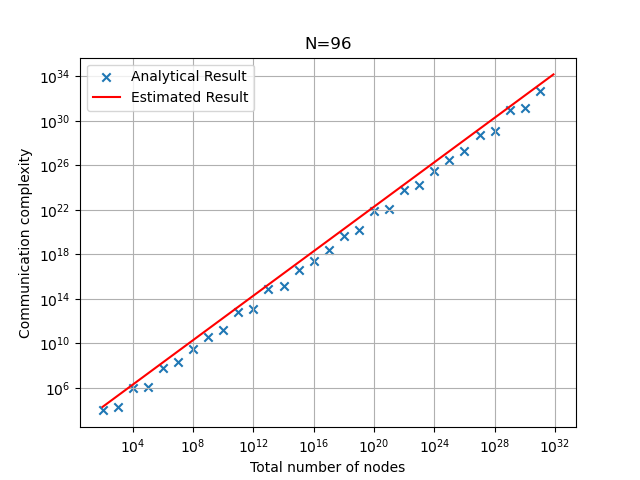}
			\caption{N=96}
			\label{fig:image4}
		\end{minipage}
		\caption{Communication Complexity Analytical Results vs Estimated Results}
	\end{figure}

The conﬁrmation delay would keep increase with increasing network depth, and is approximately equal to $mt_{ave}$, due to $m$ is the network depth. When $N$ is ﬁxed, there exists an approximate relationship between $V$ and $m$, and 
\begin{equation}
	D=mt_{ave}  \approx \frac{\log_2 V}{\log_2 2N}t_{ave}.
\end{equation}
\onecolumn
\subsection{Design of the update ordering}
In order to design the update ordering of the Hyper-simple fractal network appropriate, it is necessary to design a cyclic path along the surface of the fractal network. A feasible loop design can be provided using mathematical rotation operators. For $V_{N,m}$, any $N$ rotations $\sigma$ can be taken to use, for example,
\begin{equation}
	\sigma=(1,2,...,N),
\end{equation}
which means
\begin{equation}
	\sigma(i)=i+1,i<N;\ \sigma(N)=1.
\end{equation}

We only need to figure out which is the next node to be fresh after arbitrary $v_{\text{label}}$. It is convenient to record the next node after $v_{\text{label}}$ as $B(v_{\text{label}})$.

\textbf{1) For elements in $V_{N,1}$:}

\begin{equation}
B(v_{(i,0)})=\left\{\begin{aligned}
	&v_{(\sigma(i),0),(1,0)},&\sigma(i)\neq 1,
	\\
	&v_{(\sigma(i),0),(2,0)},&\sigma(i)=1.
\end{aligned}\right.\end{equation}

\textbf{2) For elements in $V_{N,2}-V_{N,1}$:}
\begin{equation}
    B(v_{(k,0),(k,0)})=v_{(k,0)},
\end{equation}

\begin{equation}
    B(v_{(k,0),(i,0)})=\left\{\begin{aligned}
	&v_{(k,0),(\sigma(i),0),(1,0)},&(k,0),(\sigma(i),0)\in Label_{N,2}\ and\ (k,0),(\sigma(i),0),(1,0)\in Label_{N,3},
	\\
	&v_{(k,0),(\sigma(i),0),(2,0)},&(k,0),(\sigma(i),0)\in Label_{N,2}\ and\ (k,0),(\sigma(i),0),(1,0)\notin Label_{N,3},
	\\
	&v_{(k,0),(\sigma^2(i),0),(1,0)},&(k,0),(\sigma(i),0)\notin Label_{N,2}\ and\ (k,0),(\sigma^2(i),0),(1,0)\in Label_{N,3},
	\\
	&v_{(k,0),(\sigma^2(i),0),(2,0)},&(k,0),(\sigma(i),0)\notin Label_{N,2}\ and\ (k,0),(\sigma^2(i),0),(1,0)\notin Label_{N,3}.
\end{aligned}\right.
\end{equation}

\textbf{3) For elements in $V_{N,m-1}-V_{N,2}$:}

The subscript pair label  can be written as $\text{label1},(k,0),(i,0)$ or $\text{label1},(k,1),(i,0)$.

If $\text{label1},(k,0),(i,0)\in \cup_{i=1}^{m}Label_{N,i}$ and  $\text{label1},(k,0),(\sigma(i),0)\in \cup_{i=1}^{m}Label_{N,i}$,
\begin{equation}
B(v_{\text{label1},(k,0),(i,0)})=\left\{\begin{aligned}
	&v_{\text{label1},(k,1),(\sigma(i),0),(1,0)},&if\ \text{label1},(k,0),(\sigma(i),0),(1,0)\in \cup_{i=1}^{m}Label_{N,i},
	\\
	&v_{\text{label1},(k,1),(\sigma(i),0),(2,0)},&if\ \text{label1},(k,0),(\sigma(i),0),(1,0)\notin \cup_{i=1}^{m}Label_{N,i},
\end{aligned}\right.
\end{equation}
else if $\text{label1},(k,0),(i,0)\in \cup_{i=1}^{m}Label_{N,i}$ but $\text{label1},(k,0),(\sigma(i),0)\notin \cup_{i=1}^{m}Label_{N,i}$,
\begin{equation}
B(v_{\text{label1},(k,0),(i,0)})=\left\{\begin{aligned}
	&v_{\text{label1},(k,1),(\sigma^2(i),0),(1,0)},&if\ \text{label1},(k,0),(\sigma^2(i),0),(1,0)\in \cup_{i=1}^{m}Label_{N,i},
	\\
	&v_{\text{label1},(k,1),(\sigma^2(i),0),(2,0)},&if\ \text{label1},(k,0),(\sigma^2(i),0),(1,0)\notin \cup_{i=1}^{m}Label_{N,i},
\end{aligned}\right.
\end{equation}
else
\begin{equation}
B(v_{\text{label1},(k,0),(i,0)})=\left\{\begin{aligned}
	&v_{\text{label1},(k,1),(1,0)},&if\ \text{label1},(k,0),(1,0)\in \cup_{i=1}^{m}Label_{N,i},
	\\
	&v_{\text{label1},(k,1),(2,0)},&if\ \text{label1},(k,0),(1,0)\notin \cup_{i=1}^{m}Label_{N,i}.
\end{aligned}\right.
\end{equation}

If $\text{label1},(k,1),(i,0)\in \cup_{i=1}^{m}Label_{N,i}$, and  $\text{label1},(k,1),(\sigma(i),0)\in \cup_{i=1}^{m}Label_{N,i}$,
\begin{equation}
B(v_{\text{label1},(k,1),(i,0)})=\left\{\begin{aligned}
	&v_{\text{label1},(k,1),(\sigma(i),0),(1,0)},&if\ \text{label1},(k,1),(\sigma(i),0),(1,0)\in \cup_{i=1}^{m}Label_{N,i},
	\\
	&v_{\text{label1},(k,1),(\sigma^2(i),0),(2,0)},&if\ \text{label1},(k,1),(\sigma(i),0),(1,0)\notin \cup_{i=1}^{m}Label_{N,i},
\end{aligned}\right.
\end{equation}
else if $\text{label1},(k,1),(i,0)\in \cup_{i=1}^{m}Label_{N,i}$, but  $\text{label1},(k,1),(\sigma(i),0)\notin \cup_{i=1}^{m}Label_{N,i}$,
\begin{equation}
B(v_{\text{label1},(k,1),(i,0)})=\left\{\begin{aligned}
	&v_{\text{label1},(k,1),(\sigma^2(i),0),(1,0)},&if\ \text{label1},(k,1),(\sigma^2(i),0),(1,0)\in \cup_{i=1}^{m}Label_{N,i},
	\\
	&v_{\text{label1},(k,1),(\sigma^2(i),0),(2,0)},&if\ \text{label1},(k,1),(\sigma^2(i),0),(1,0)\notin \cup_{i=1}^{m}Label_{N,i},
\end{aligned}\right.\end{equation}
else
\begin{equation}B(v_{\text{label1},(k,1),(i,0)})=\left\{\begin{aligned}
	&v_{\text{label1},(k,0)},&if\ k\ne 1 \ or\ N,
	\\
	&v_{\text{label1},(k_0,0)},&else,
\end{aligned}\right.\end{equation}
where $k_0$ satisfies $\text{label1},(k_0,0)\notin \cup_{i=1}^mLabel_{N,i}$.

\textbf{4) For elements in $V_{N,m}-V_{N,m-1}$:}

The subscript pair label can be written as $\text{label1},(k,0),(i,0)$ or $\text{label1},(k,1),(i,0)$. 

When label can be written as $\text{label1},(k,0),(i,0)$, 

	\begin{equation}
	B(v_{\text{label1},(k,0),(i,0)})=\left\{\begin{aligned}
		&v_{\text{label1},(k,0),(\sigma(i),0)},& i\ne  N,
		\\
		&v_{\text{label1},(k,1),(1,0)},& i= N.
	\end{aligned}\right.
	\end{equation}
 
When label can be written as $\text{label1},(k,1),(i,0)$,
	\begin{equation}
	B(v_{\text{label1},(k,1),(i,0)})=\left\{\begin{aligned}
		&v_{\text{label1},(k,1),(\sigma(i),0)},& i\ne  N,
		\\
		&v_{\text{label1},(k,0)},& i=  N, k\ne 1 \ or \ N,
		\\
		&v_{\text{label1},(k_0,0)},& i=  N,k= 1 \ or\ N,
	\end{aligned}\right.
	\end{equation}
	where $k_0$ satisfies $\text{label1},(k_0,0)\notin \cup_{i=1}^mLabel_{N,i}$.

 \twocolumn

\section{CONCLUSION}

Our work introduces a novel mathematical framework that unifies fractal geometry, topology, networking science, blockchain data structure, and distributed consensus to revolutionize blockchain network design. This approach not only solves critical challenges in blockchain scalability and efficiency but also provides a powerful toolkit for understanding adaptive network behaviors across diverse fields. By bridging abstract mathematics with practical engineering, we demonstrate how interdisciplinary research can lead to breakthrough solutions in network science.

Future work will focus on refining these models for broader applications in social and urban networks, potentially transforming our approach to complex systems design and analysis. This research opens new avenues for exploring the fundamental principles governing large-scale, dynamic networks of blockchain, promising significant advancements in both theoretical mathematics and practical network engineering.

\section{ACKNOWLEDGMENT}
The author list for this paper is currently tentative. We are in communication with all potential authors to determine the final author list and order. We appreciate all individuals who have contributed to this research, and the final author list will be confirmed upon formal publication.

Funding support:
Fundamental Research Funds for the Central Universities, Tongji University, Project No. 22120240564.

\bibliographystyle{IEEEtran}
\bibliography{reference}
\end{document}